\documentclass{appolb}
\usepackage{graphicx}

\usepackage{amsmath}
\usepackage{amssymb}
\usepackage{amsfonts}
\usepackage{graphicx}
\usepackage{pstricks}
\usepackage{tikz}
\usepackage{mathtools}

\newcommand{\vnabla}{\boldsymbol{\mathbf\nabla}}

\newcommand{\bJ}{\boldsymbol{\mathbf J}}

\newcommand{\nn}{\nonumber}

\newcommand{\be}{\begin{equation}}
\newcommand{\ee}{\end{equation}}
\newcommand{\bqr}{\begin{eqnarray}}
\newcommand{\eqr}{\end{eqnarray}}

\begin{document}

\title{Skyrme N2LO pseudo-potential for calculations of properties of atomic nuclei}

%\author{P. Becker, D. Davesne, J. Meyer 
%\address{Universit\'e de Lyon, Universit\'e Lyon 1, 43 Bd. du 11 Novembre 1918, F-69622 Villeurbanne cedex, France\\
%              CNRS-IN2P3, UMR 5822, Institut de Physique Nucl{\'e}aire de Lyon \\}
%\mbox{} \\
%J. Navarro
%\address{IFIC (CSIC-Universidad de Valencia), Apartado Postal 22085, E-46.071-Valencia, Spain \\}
%\mbox{}\\
%A.Pastore
%\address{Department of Physics, University of York, Heslington, York, Y010 5DD, United Kingdom}
%
%}
\author{P. Becker$^{a,b}$, D. Davesne$^{a,b}$, J. Meyer$^{a,b}$, J. Navarro$^{c}$, A. Pastore$^{d}$
\address{$^{a}$ Universit\'e de Lyon, Universit\'e Lyon 1, 43 Bd. du 11 Novembre 1918, F-69622 Villeurbanne cedex, France\\
$^{b}$              CNRS-IN2P3, UMR 5822, Institut de Physique Nucl{\'e}aire de Lyon \\
$^{c}$ IFIC (CSIC-Universidad de Valencia), Apartado Postal 22085, E-46.071-Valencia, Spain \\
$^{d}$ Department of Physics, University of York, Heslington, York, Y010 5DD, United Kingdom}
}
\maketitle

%%%%%%%%%%%%%%%%%%%%%%%%%%%%%%%%%%%%%%

\begin{abstract}
We present recent developments obtained in the so-called N2LO extension of the usual Skyrme pseudo-potential. In particular, we 
%extend the discussion made in Ref.~\cite{bec17b} by inspecting 
discuss the isovector splitting mass in infinite nuclear matter and the pairing gaps of selected semi-magic even-even nuclei. 
\end{abstract}
\PACS{  21.30.Fe       % Forces in hadronic systems and effective interactions
             21.60.Jz        % Nuclear Density Functional Theory and extensions
                                   % (includes Hartree-Fock and random-phase approximations)
             21.65.-f          % Nuclear matter
             21.65.Mn}
  
%%%%%%%%%%%%%%%%%%%%%%%%%%%%%%%%%%%%%%

\section{Introduction}

%%%%%%%%%%%%%%%%%%%%%%%%%%%%%%%%%%%%%%

Skyrme's original idea~\cite{sky59} was to build an effective zero-range pseudo-potential as a momentum expansion of a given finite-range form factor.
In Ref.~\cite{dav16ANNALS}, we have explicitly discussed how it is possible to derive (starting from a finite-range interaction such as Gogny~\cite{dec80} or M3Y~\cite{nak03}), all the terms of the Skyrme interaction up to any order. At order three, the results are in agreement with the previous calculations done in Refs.~\cite{car08,rai11} when one imposes that the N3LO\footnote{N$\ell$LO with $\ell=1,2,3,\dots$ is the name given to the pseudo-potential or a functional when the highest power of the momentum operator kept before truncating the expansion is $k^{2\ell}$.} pseudo-potential is invariant under  Galilean and local gauge transformations.

The main motivations behind exploring such \emph{extended} versions of the Skyrme pseudo-potential  have been discussed in Ref.~\cite{kor14}: the current discrepancies observed between predicted values with the \emph{standard} Skyrme pseudo-potential and the measured observables can't be further reduced by using improved fitting procedures. It is thus time to explore richer functionals that may help us getting theoretical predictions closer to the experimental measurements.

The first decisive step can be found in Ref.~\cite{car10} where the authors have used for the very first time the extended N3LO functional~\cite{car08} to test the role of higher order terms. By means of density matrix expansion (DME), they have shown that the next-to-next-to-leading order (N2LO) plays an important role in reducing by roughly one order of magnitude the discrepancy between the exact result and the DME expansion. 
By extending the formalism to N3LO, the DME results further improve and get closer and closer to the \emph{exact} values, thus showing that the expansion converges.

Then, in Ref.~\cite{bec17b}, we have performed the very first study of the extended Skyrme N2LO pseudo-potential in the case of spherical even-even nuclei. In that article, we have obtained the first parametrisation of such a pseudo-potential using properties of some selected double-magic nuclei.
In the present article, we continue our investigation, by extending our analysis to open-shell nuclei and in particular exploring the behaviour of pairing gaps along some isotopic chains~\cite{bri05}  using the N2LO pseudo-potential. 

The article is organised as follows: in Sec.~\ref{sec:n2lo} we briefly summarise the key-concepts of the N2LO Skyrme pseudo-potential. In Sec.\ref{eff:mass}, we discuss the properties of the effective mass and in  Sec.~\ref{sec:err} we discuss pairing properties of the SN2LO1 interaction. We finally provide our conclusions in Sec.~\ref{sec:conc}.

%%%%%%%%%%%%%%%%%%%%%%%%%%%%%%%%%%%%%%

\section{Skyrme N2LO}\label{sec:n2lo}

%%%%%%%%%%%%%%%%%%%%%%%%%%%%%%%%%%%%%%

The N2LO Skyrme pseudo-potential, as described in Refs.~\cite{car08,rai11,dav13b}, is a generalisation of the standard Skyrme interaction, corresponding to the expansion of the momentum space matrix elements of a generic interaction in powers of the relative momenta $\mathbf{k}, \mathbf{k}'$ up to the fourth order. It is written as the sum of three terms~\cite{dav14c}
\begin{equation}\label{eq:N2LO}
V_{\text{N2LO}} =V_{\rm N2LO}^{C}+V_{\rm N1LO}^{LS}+V_{\rm N1LO}^{DD}\;.%+V_{\rm N2LO}^{T} .
\end{equation}
The central term reads
\begin{eqnarray} \label{eq:N2LO:c}
V_{\rm N2LO}^{C} &=& t_0 (1+x_0 P_{\sigma}) + \frac{1}{2} t_1 (1+x_1 P_{\sigma}) ({\mathbf{k}}^2 + {\mathbf k'}^2)   + t_2 (1+x_2 P_{\sigma}) ({\mathbf k} \cdot {\mathbf k'})  \nonumber\\
            & & + \frac{1}{4} t_1^{(4)} (1+x_1^{(4)} P_{\sigma}) \left[({\mathbf k}^2 + {\mathbf k'}^2)^2 + 4 ({\mathbf k'} \cdot {\mathbf k})^2\right] \nn\\
            &&+ t_2^{(4)} (1+x_2^{(4)} P_{\sigma}) ({\mathbf k'} \cdot {\mathbf k}) ({\mathbf k}^2 + {\mathbf k'}^2) .
\end{eqnarray}
In the above expression, a Dirac function $\delta({\mathbf r}_1-{\mathbf r}_2)$
is to be understood~\cite{ben03}. The density-dependent term $V_{\rm N1LO}^{DD}$ and the spin-orbit term $V_{\rm N1LO}^{LS}$ have the same structure as in the standard Skyrme interaction~\cite{cha97}. An alternative to the use of a density-dependent term would be the inclusion of an explicit three-body term. This possibility has been discussed in details in Ref.~\cite{sad13}.

From the interaction given in Eq.~(\ref{eq:N2LO}), we are in position to derive the corresponding functional form by averaging over Hartree-Fock (HF) states. Since the focus of the article is the study of semi-magic even-even nuclei, we limit ourselves to the time-even spherically-symmetric case and we obtain
\bqr
\label{eq:EDF_N1LO_C_sphere}
\mathcal{E}&=&\sum_{t=0,1}   C^{\rho}_t \, \rho_t^2+C^{\Delta \rho}_t  \, \rho_t \Delta \rho_t  +    C^{\tau}_t  \, \rho_t \tau_t\, - \,   \tfrac{1}{2}\,  C^T_t  \, J_t^2+  C^{\nabla J}_t \, \rho_t \, \vnabla \cdot \bJ_t                 \nn\\
  &+& C^{(4) \Delta \rho}_t \, \left( \Delta \rho_t \right)^2  +  C^{(4) M \rho}_t \Big(   \rho_t Q_t  \, + \, \tau_t^2 \,  \Big) \nn \\
  & +&  C^{(4) M \rho}_t \, \Big[\,  \mbox{Re}(\tau_{t, \mu \nu}) \mbox{Re}(\tau_{t, \mu \nu}) 
 \, - \mbox{Re}(\tau_{t, \mu \nu}) \nabla_{\mu} \nabla_{\nu} \rho_t  \, \Big]              \nn \\
  & -  &  C^{(4) M s}_t
                      \, \left[ \, \left( \nabla_{\mu} J_{t, \mu \nu} \right)^2
 \, + \, 4 J_{t, \mu \nu} V_{t, \mu \nu} - \mbox{Im}(K_{t,\mu \nu \kappa}) \mbox{Im}(K_{t,\mu \nu \kappa}) \, \right]     \,.
\end{eqnarray}
\noindent We refer to Ref.~\cite{bec17b} for a more detailed discussion on the properties of the N2LO functional.
\noindent By comparing Eq.~(\ref{eq:EDF_N1LO_C_sphere}) with the standard form of the Skyrme functional (N1LO), see  for example Ref.~\cite{les07}, we observe the appearance of new densities: $V_{t,\mu\nu},K_{t,\mu\nu\kappa},Q_t,\tau_{t,\mu\nu}$ whose complete expressions can be found in Ref.~\cite{bec17b}. In this reference, we have also given some other expressions of scalar quantities related to these new densities. 

%%%%%%%%%%%%%%%%%%%%%%%%%%%%%%
\begin{figure}[h!]
\centerline{%
\includegraphics[width=0.7\textwidth]{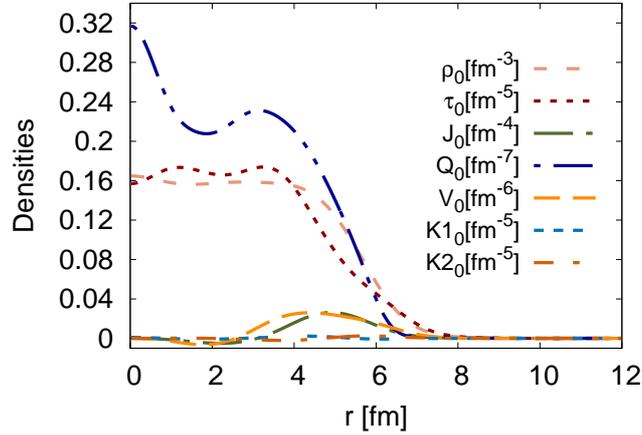}
}
\caption{(Colors online) Isoscalar densities in  $^{132}$Sn obtained with SN2LO1 functional. See text for details.}
\label{Fig:densities}
\end{figure}
%%%%%%%%%%%%%%%%%%%%%%%%%%%%%%

In order to have a physical insight and give some order of magnitude, we show in Fig.~\ref{Fig:densities}, the radial profile of all these local densities for $^{132}$Sn using the SN2LO1 interaction.
We observe that the density $Q$ has similar order of magnitude and shape as the kinetic density $\tau$, while the other additional densities are  peaked at the surface of the nucleus and almost zero in the bulk of the nucleus. It is presently difficult to estimate the respective role of these new densities since only one parametrisation is available, but we plan to investigate in the near future their properties in a more systematic way along the nuclear chart. 

%{\bf DD : attention : les densit\'es comportaient initialement des indices 0 dans le texte mais pas sur la figure. A voir...}

%%%%%%%%%%%%%%%%%%%%%%%%%%%%%%%

\section{Effective mass}\label{eff:mass}

%%%%%%%%%%%%%%%%%%%%%%%%%%%%%%%

\begin{figure}[htb]
\centerline{%
\includegraphics[angle=-90,width=0.55\textwidth]{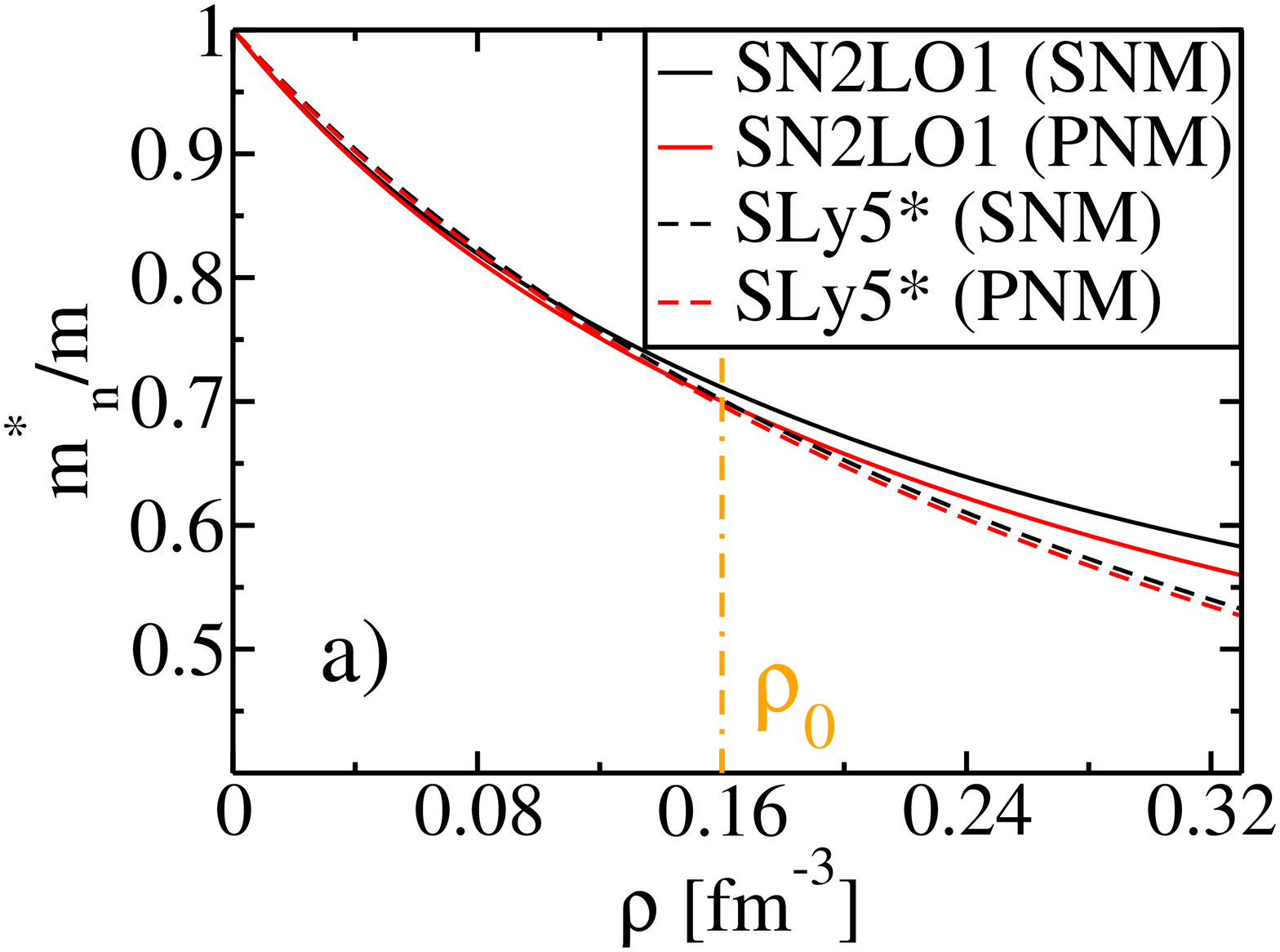}
\includegraphics[angle=-90,width=0.55\textwidth]{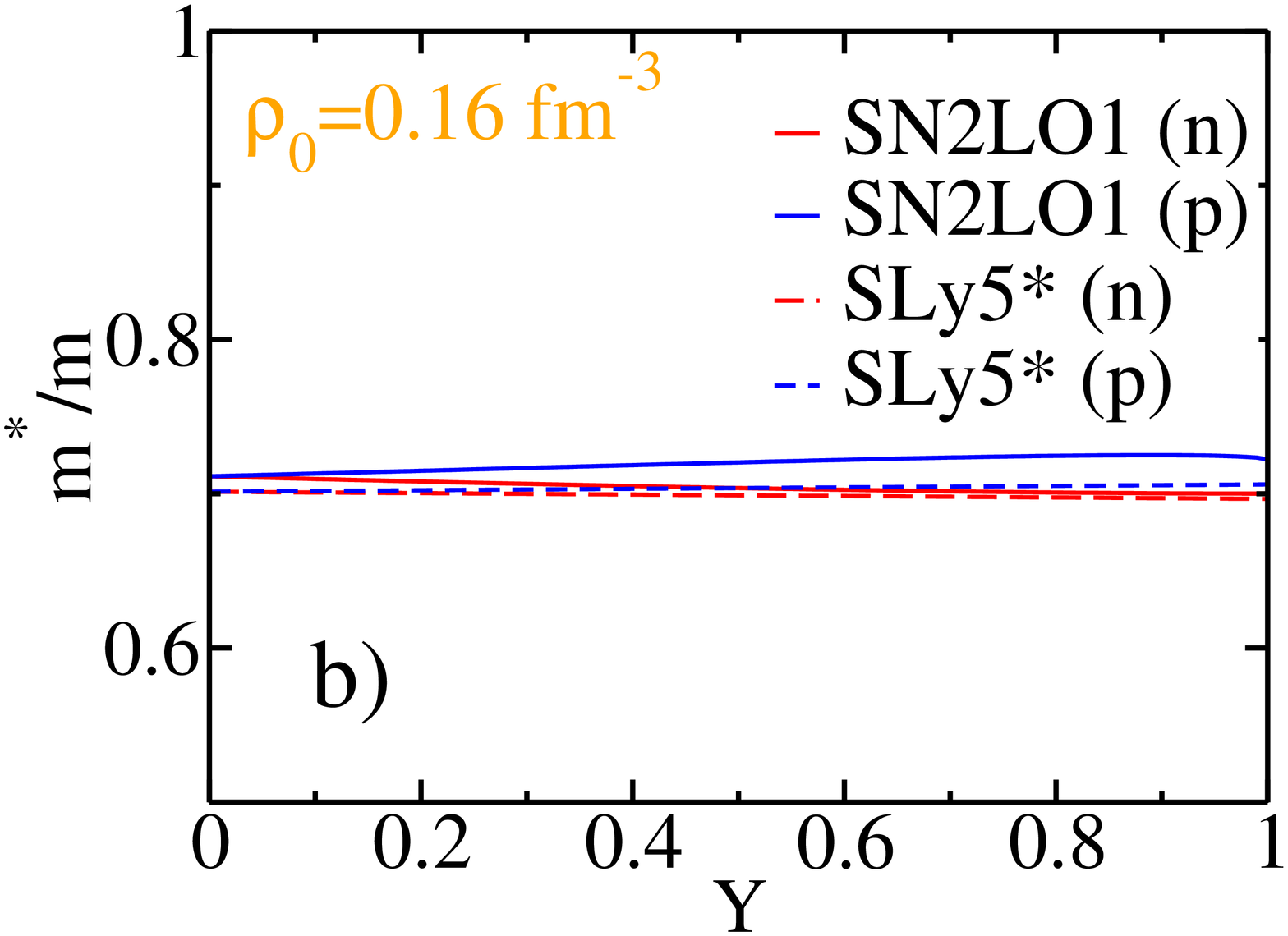}
\vspace{2mm}
}
\caption{(Colors online) Panel a: neutron effective mass calculated in SNM and PNM for the two interactions as a function of the density of the infinite system. The vertical dashed line represents the saturation density $\rho_0$ of SNM. See text for details. In Panel b we show the evolution of neutron and proton effective masses at saturation density and in function of the asymmetry parameter Y.}
\label{fig:mass}
\end{figure}

The value of the effective mass at saturation density~\cite{les06} has a strong impact on the density of states around the Fermi energy and thus on the underlying pairing properties. In the fitting protocol used to adjust SN2LO1, we have imposed  an explicit constraint on the value of the effective mass $m^*/m$ at saturation density ($\rho_0$=0.16 fm$^{-3}$) in infinite symmetric nuclear matter (SNM). From our fit, we have obtained $m^*/m=0.71$; such a value should be compared with the one derived for SLy5*    $m^*/m=0.7$ and fitted with a similar protocol~\cite{pas13}. In Fig.~\ref{fig:mass} a, we show the evolution of the effective neutron mass, $m^*_n/m$, as a function of the density $\rho$ for SNM and pure neutron matter (PNM) for both interactions.

%%%%%%%%%%%%%%%%%%%%%%%%%%%%%%%
%%%%%%%%%%%%%%%%%%%%%%%%%%%%%%%

\noindent We observe that  in both cases (SLy5* and SN2LO1), the behaviour of neutron 
 effective mass is very similar. In both cases, we observe the absence of poles in the effective mass up to very high densities. 
 In  Fig.~\ref{fig:mass} b, we show the evolution of proton and neutron effective masses as a function of the asymmetry parameter Y~\cite{Dav15AA}, from SNM (Y=0) to  PNM (Y=1). We notice that  mass splitting is essentially zero at saturation density. The splitting is slightly bigger in SN2LO1 than in SLy5*, but they are very close to zero. 
%The value we obtain here is not yet in agreement with \emph{ab-initio} results
We refer the reader to Ref.~\cite{les06} for a devoted study of the isovector mass splitting in Skyrme functionals. 
%

%%%%%%%%%%%%%%%%%%%%%%%%%%%%%%%

\section{Pairing gaps}\label{sec:err}

%%%%%%%%%%%%%%%%%%%%%%%%%%%%%%%

\begin{figure}[htb]
\centerline{%
\includegraphics[angle=-90,width=0.55\textwidth]{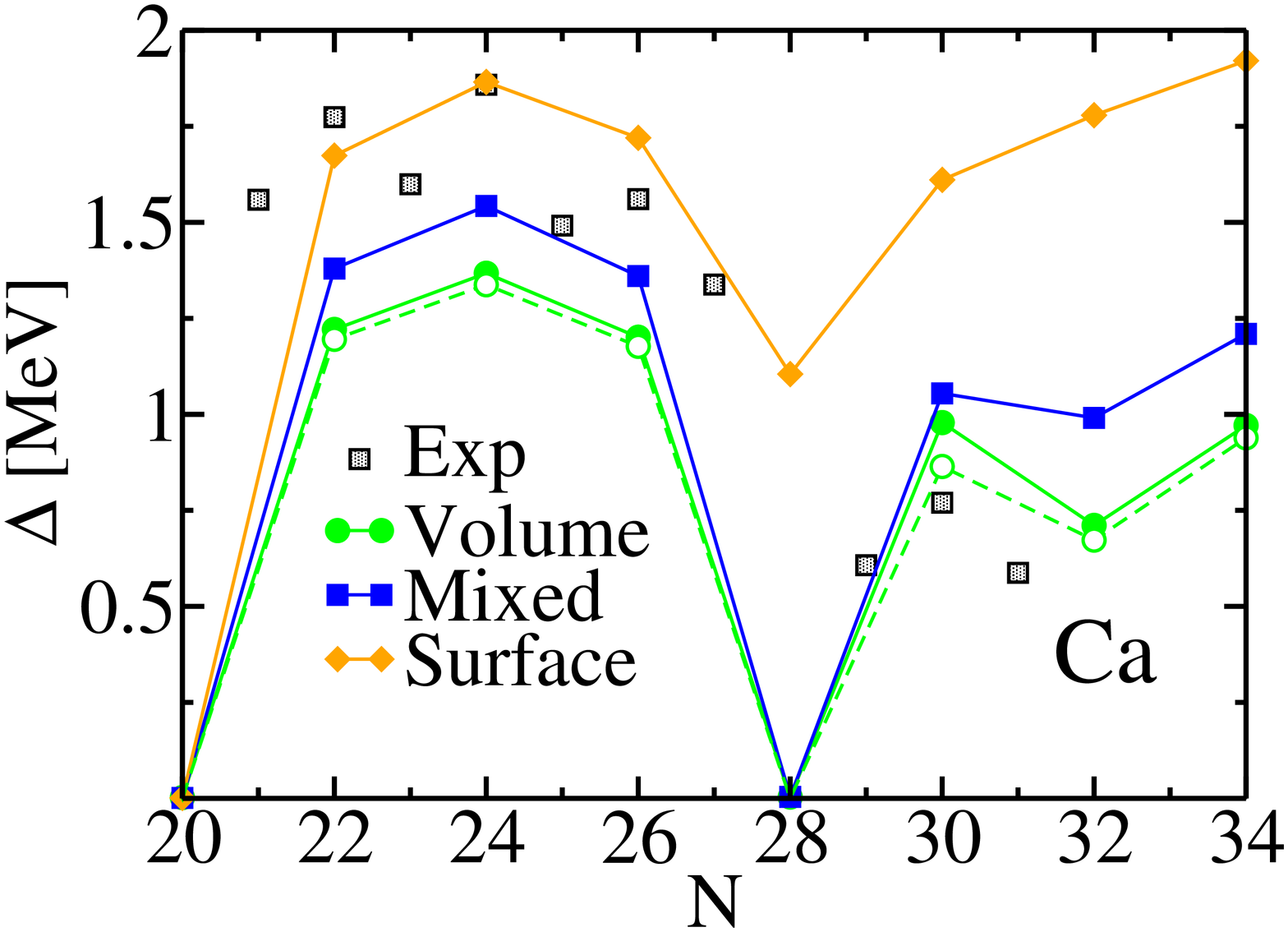}
\includegraphics[angle=-90,width=0.55\textwidth]{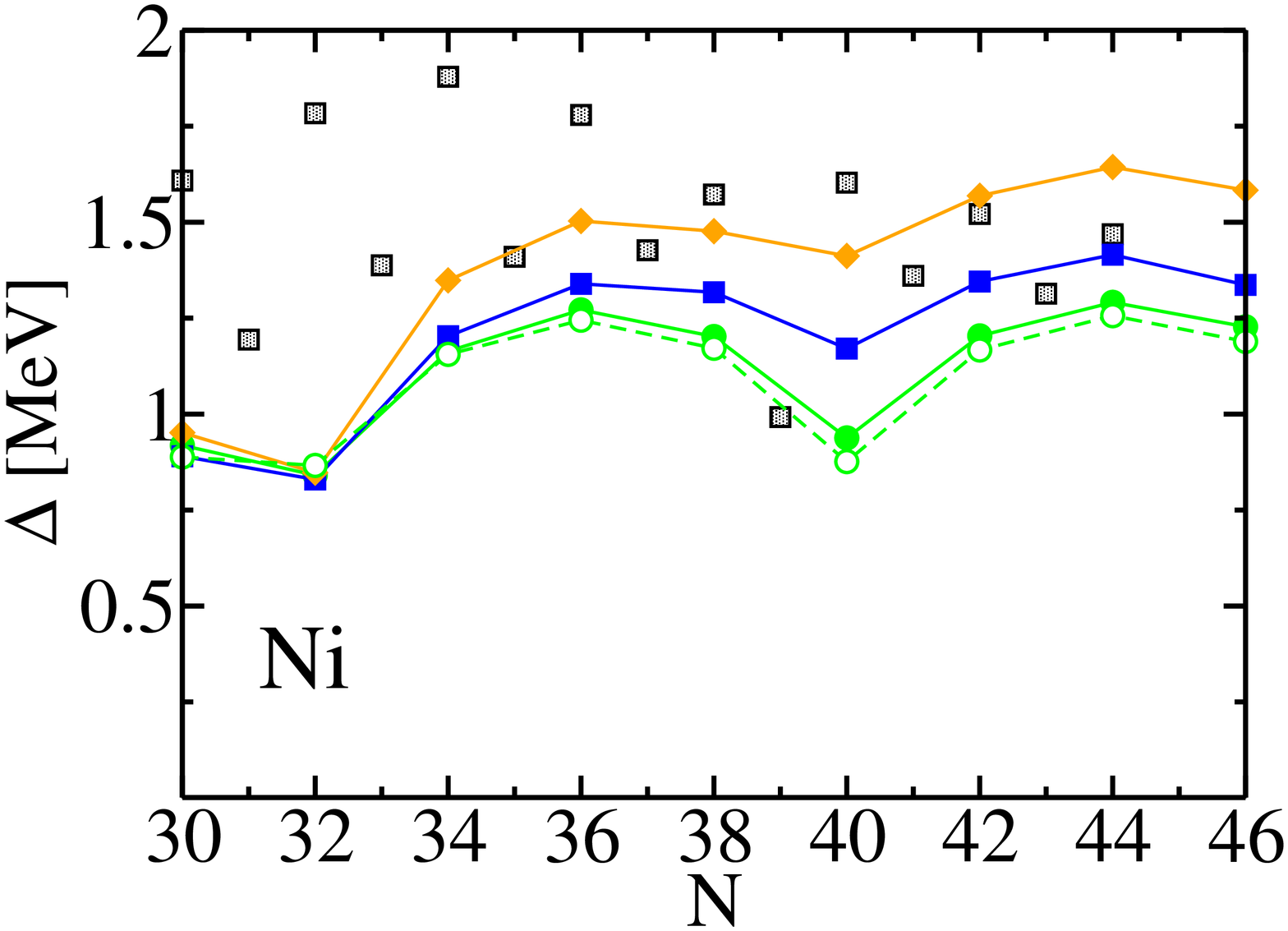}
}
\centerline{%
\includegraphics[angle=-90,width=0.55\textwidth]{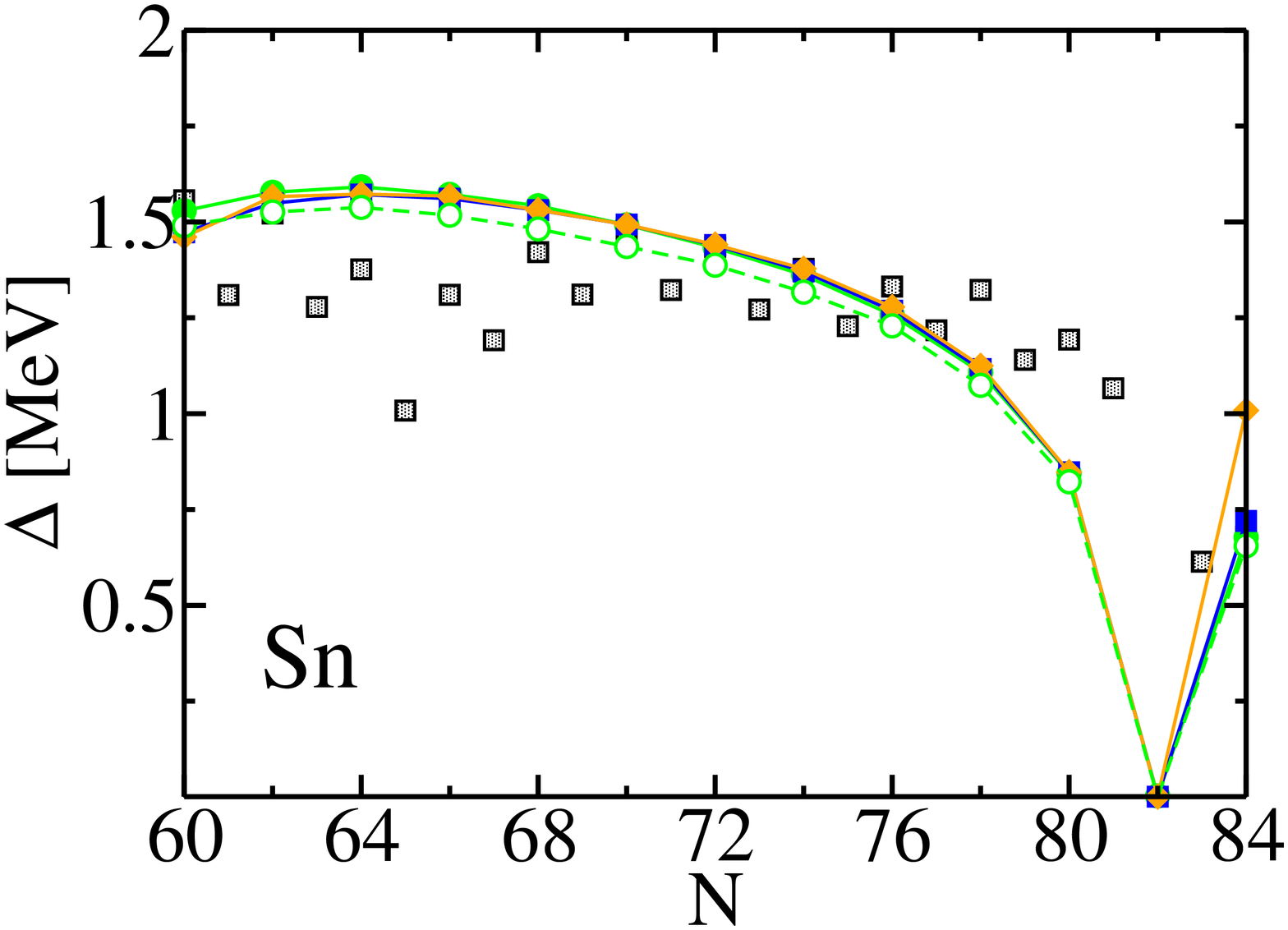}
\includegraphics[angle=-90,width=0.55\textwidth]{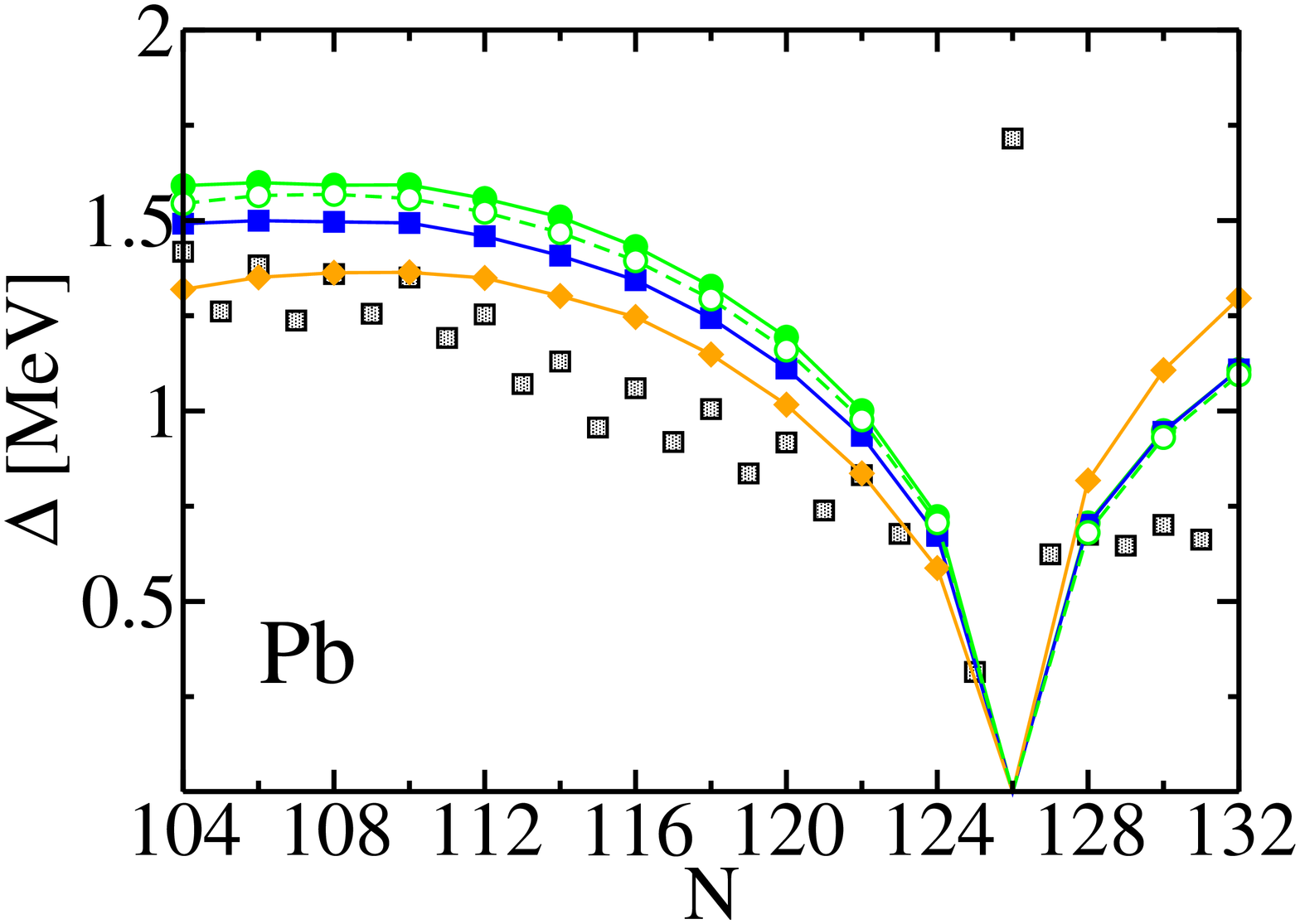}
}
\caption{(Colors online) Evolution of average pairing gaps for Ca, Ni, Sn and Pb isotopic chains in function of the neutron number $N$. Full symbols correspond to the SN2LO1 interaction, while the empty ones to SLy5*.The theoretical pairing gaps have been calculated according to Eq.\ref{eq:gap}. See text for details.}
\label{Fig:gap}
\end{figure}

In this section, we perform a first systematic study of the pairing gaps using the SN2LO1 functional. As for the SLy5* functional~\cite{pas13}, it has been fitted using properties of infinite nuclear matter and doubly-magic nuclei with the additional stability constraint coming from Linear Response theory~\cite{hel13,Pas15T,report}. This particular choice leaves us complete freedom in determining the parameters entering the pairing channel. Therefore, for the current analysis, we have decided to use the numerical code \verb|WHISKY|~\cite{bec17b} to solve the Hartree-Fock-Bogoliubov equations (HFB)~\cite{rin04} for a simple density dependent pairing interaction of the form~\cite{ber91b}
\begin{eqnarray}
V_{pair}(\mathbf{r}_1,\mathbf{r}_2)=V_0\left[ 1-\eta \frac{\rho(\mathbf{R})}{\rho_0}\right]\delta(\mathbf{r}_1-\mathbf{r}_2)\;,
\end{eqnarray}
where $\mathbf{R}=(\mathbf{r}_1+\mathbf{r}_2)/2$ is the center of mass of the interacting particles and $V_0$ is the strength of the interaction. To avoid an ultraviolet divergency associated with such a contact interaction~\cite{bul02}, we adopted a sharp-cut off in the quasi-particle space $E_{cut}=60$ MeV. It is then possible to determine the pairing field $\Delta(r)$. As it is well-known, its shape can be modified  by setting the value of the parameter $\eta$ to $0,1/2$ and $1$, thus producing a volume, mixed and surface pairing field~\cite{san05}, respectively. To quantify the amount of pairing correlations, we calculated the average pairing gap defined as~\cite{pas13p}
\begin{eqnarray}\label{eq:gap}
\bar{\Delta}=\frac{\int  \Delta(\mathbf{r})\tilde{\rho}(\mathbf{r})d\mathbf{r}}{\int \tilde{\rho}(\mathbf{r})d\mathbf{r}}\;,
\end{eqnarray}
where $\tilde{\rho}_q$ is the pairing density. The results are depicted in Fig.~\ref{Fig:gap}. The average pairing gaps have been obtained using SN2LO1 for the three values of $\eta = 0,1/2,1$ mentioned above as a function of the neutron number $N$. The experimental values have been  extracted from experimental binding energies~\cite{aud03} using the standard the three-point formula~\cite{ben00}. Since our numerical code works in spherical symmetry only, we limited ourselves to the four isotopic chains of semi-magic nuclei.

%%%%%%%%%%%%%%%%%%%%%%%%%%%%%%%
%%%%%%%%%%%%%%%%%%%%%%%%%%%%%%%

\noindent For these calculations, we have used a pairing strength  of $V_0=-200$ MeV \; fm$^{3}$ (volume),  $V_0=-300.2$ MeV\;fm$^{3}$ (mixed) and  $V_0=-496.6$ MeV\;fm$^{3}$ (surface). Since the pairing interaction is not adjusted during the fitting procedure, we adjusted in a completely arbitrary way the pairing strength to the same value of pairing gap in $^{120}$Sn. From this figure we can conclude that the underlying single-particle spectrum is reasonable and the resulting pairing gaps give a reasonable description of available experimental data. For completeness, we have also repeated the same calculations using SLy5* for the mean field and the same pairing interaction. In the figure, we have reported only (for clarity) the case of volume-type pairing interaction (same strength). The two other not reported cases give similar results compared to the SN2LO1 ones. The small differences observed among the two interactions are due to small differences in the underlying single-particle structure.

%%%%%%%%%%%%%%%%%%%%%%%%%%%%%%%

\section{Conclusions}\label{sec:conc}

%%%%%%%%%%%%%%%%%%%%%%%%%%%%%%%

In this article, we have discussed some important properties of the extended Skyrme interaction N2LO. By using the set of parameters determined in Ref.~\cite{bec17b}, we have first studied the behaviour of the effective mass as a function of the density and isospin asymmetry of the infinite nuclear medium. The isovector mass splitting is still not compatible with $ab-initio$ findings~\cite{bom91}, but the current results go in the right direction. Therefore, one may expect, adding such explicit constraint into the fitting protocol, that we should be able to obtain a positive splitting~\cite{Dav15AA}.

We have  discussed the isotopic evolution of pairing gaps for some relevant isotopic chains of semi-magic nuclei using the SN2LO1 functional plus density-dependent contact pairing interaction. At present the extended functional provides us with results that are qualitatively of the same level of accuracy as SLy5*, both being fitted with a very similar fitting protocol including the explicit  constraint on the linear response of infinite nuclear matter~\cite{report}. This is an improvement compared to the vast majority of existing functionals that manifest instabilities in the spin-channels. Unfortunately at present, we have not been able to find a quantitative indicator that proves the necessity of using the extended pseudo-potential compared to the simpler SLy5*. A more rigorous statistical analysis of the properties of the new functional is now mandatory~\cite{dob14} as well as a rethinking of the penalty function used during the fit to identify the most relevant observables that may help better constraining higher order parameters and avoid sloppiness~\cite{nikvsic2016sloppy} in some directions of parameter space.

%%%%%%%%%%%%%%%%%%%%%%%%%%%%%%%

\section*{Acknowledgments}

%%%%%%%%%%%%%%%%%%%%%%%%%%%%%%%

The work of J.N. has been supported by grant FIS2014-51948-C2-1-P, Mineco (Spain).
The work of A.P. is supported  by the UK Science and Technology Facilities Council under Grants No. ST/L005727 and ST/M006433. 

% [LIMIT 8 PAGES]
\bibliographystyle{polonica}
\bibliography{biblio}

\end{document}